\documentclass[aps,pre,reprint,superscriptaddress]{revtex4-2}
\usepackage{siunitx}
\usepackage[dvipdfmx]{graphicx}
\usepackage{bm}
\usepackage{url}
\usepackage{amsmath,amssymb}
\usepackage{color}
\usepackage{comment}

\newcommand{\cbk}[1]{{\left( #1 \right)}}
\newcommand{\p}{\partial}
\newcommand{\pd}[2]{\cfrac{\p #1}{\p #2}}

\newcommand{\od}[2]{\cfrac{d #1}{d #2}}
\newcommand{\odt}[2]{\cfrac{d^2 #1}{{d #2}^2}}

\begin{document}

\title{Pairing-induced motion of source and inert particles driven by surface tension}

\author{Hiroaki Ishikawa}
\affiliation{Department of Physics, Chiba University, 1-33 Yayoi-cho, Inage-ku, Chiba 263-8522, Japan}

\author{Yuki Koyano}
\affiliation{Department of Physics, Graduate School of Science, Tohoku University, 6-3, Aoba, Aramaki, Aoba-ku, Sendai 980-8578, Japan }
\thanks{present address : Graduate School of Human Development and Environment, Kobe University, 3-11 Tsurukabuto, Nada-ku, Kobe, Hyogo 657-0011, Japan}

\author{Hiroyuki Kitahata}
\affiliation{Department of Physics, Chiba University, 1-33 Yayoi-cho, Inage-ku, Chiba 263-8522, Japan}

\author{Yutaka Sumino}
\email{ysumino@rs.tus.ac.jp}
\affiliation{Department of Applied Physics, Faculty of Science Division I, Tokyo University of Science, 6-3-1 Nijuku, Katsushika-ku, Tokyo 125-8585, Japan}
\affiliation{WaTUS and DCIS, Research Institute for Science \& Technology, Tokyo University of Science, 6-3-1 Nijuku, Katsushika-ku, Tokyo 125-8585, Japan}

\date{\today}

\begin{abstract}
We experimentally and theoretically investigate systems with a pair of source and inert particles that interacts through the concentration field. The experimental system comprises a camphor disk as the source particle and a metal washer as the inert particle. Both are floated on a red aqueous solution at various concentrations, where the glycerol modifies the viscosity of the aqueous phase. The particles form a pair owing to the attractive lateral capillary force. As the camphor disk spreads surface-active molecules at the aqueous surface, the camphor disk and metal washer move together, driven by the surface tension gradient. The washer is situated in the front of the camphor disk, keeping the distance constant during their motion, which we call a pairing-induced motion. The pairing-induced motion exhibited a transition between circular and straight motions as the glycerol concentration in the aqueous phase changed. Numerical calculations using a model that considers forces caused by the surface tension gradient and lateral capillary interaction reproduced the observed transition in the pairing-induced motion. Moreover, this transition agrees with the result of the linear stability analysis on the reduced dynamical system obtained by the expansion with respect to the particle velocity. Our results reveal that the effect of the particle velocity cannot be overlooked to describe the interaction through the concentration field.
\end{abstract}

\maketitle

\section{Introduction}

In nonequilibrium systems, a particle can spontaneously move by consuming free energy, and this is known as a self-propelled particle. The mechanism of such a self-propelled motion of a single particle has been intensively studied ~\cite{Colle_Surf_of_Chem_Act_Particles, vandadi_kang_masoud_2017, Colloid_Trans_Interfacial_F,water_droplet,chemical_reaction_droplet}, in addition to their collective behavior ~\cite{Ramaswamy_review,Ramaswamy_review_2,spontaneous_motion_iso_particles,RevModPhys_active_particles_in_compand_crow,vicsek2012,chate}. Recently, self-propelled particles that interact through a concentration field have attracted attentions as an analog for chemotactic motions of living organisms, e.g., $\it{Dictyostelium}$~\cite{dictyostelium} and $\it{E.~coli}$~\cite{ecoli}. {In an actual biological situation, the system of interest possesses macroscopic dynamics due to several species interacting within the system~\cite{Chase-and-run}. As a pioneering theoretical work, Canalejo et al. reported the active phase separation in the multiparticle system of binary species~\cite{golestanian}. In their model, the dynamics of concentration field is adiabatically eliminated. Thus, the particles interact through the concentration fields in an instantaneous manner. However, the concentration field can have their own spatio-temporal dynamics, such as diffusion and chemical reaction, which may alter the characteristics of the particle motion. In fact, the concentration field and the particle position are introduced as independent variables in the model for the camphor disk motion on water~\cite{nagayama, hirose, hayashima, koyano}. This model can show the self-propulsion through spontaneous symmetry breaking owing to the dynamics of the concentration field. Therefore, we investigate the systems with two types of particles coupled with the concentration field having their own dynamics. In the present study, we focus on a simple system that comprises a pair of source and inert particles, both of which are driven by a common concentration field.

In our experiment, we use a camphor disk and metal washer as the source and inert particles, respectively, at the surface of a glycerol aqueous solution. From the camphor disk floating at the surface of the aqueous solution, camphor molecules are continuously spread to the aqueous surface, and then, they are sublimated to the air phase. A consequent spatial gradient of surface tension drives the camphor disk and metal washer~\cite{Tomlinson, Boniface, Rotator, 3Dprinter, Grzybowski_1, Grzybowski_2, nishimori, Oliver, pccp}.  The floating metal washer distorts the surface, which causes the attractive lateral capillary force between the camphor disk and metal washer~\cite{l_cap_f}. Owing to the surface tension gradient and attractive capillary force, the camphor disk and metal washer exhibit the motion with a constant mutual distance, a pairing-induced motion. The glycerol concentration in the aqueous phase was varied as a control parameter, which changes the viscosity of the aqueous phase. We observed the transition between circular and straight motions as the glycerol concentration in the aqueous phase changed. Moreover, we constructed a mathematical model consisting of the reaction-diffusion equation for the surface-active molecules and equations of motion for the source and inert particles. Numerical results exhibit the transition from the straight to circular pairing-induced motion. This transition is analytically explained based on the bifurcation theory.

\section{Experiments}

Camphor and glycerol were obtained from Fuji-film Wako Pure Chemical (Osaka, Japan). Water was purified with the Millipore Milli-Q system (Merck, Darmstadt, Germany). Two kinds of stainless washers were obtained from Ohsato (Tokyo, Japan). One is used to be embedded in a camphor disk as a source particle, and the other is used as an inert particle.  We prepared a camphor disk (diameter \SI{7}{mm}, height \SI{2}{mm}, weight \SI{0.11}{g}) embedding a stainless washer (diameter \SI{6}{mm}, height \SI{0.4}{mm}, weight \SI{0.05}{g}) using a pellet press die set and a compression molding (Pike technologies, Madison, USA). \SI{400}{ml} of glycerol aqueous solution with the concentration of $C$ or pure water was poured into a square-shaped polystyrene container (245~mm $\times$ 245~mm $\times$ 25~mm). We put a metal washer (diameter \SI{7.5}{mm}, height \SI{0.5}{mm}, weight \SI{0.13}{g}) at the glycerol aqueous surface and then put the camphor disk. The weight of the metal washer was appropriately chosen to enhance the effect of the lateral capillary force. The shadows of the camphor disk and metal washer at the aqueous surface were recorded from below using a digital CMOS video camera (DMK37BUX273, The Imaging Source, Bremen, Germany) at 30 fps. The experiments were conducted at 25 $\pm$ 3~$^\circ$C, and the captured images were analyzed using ImageJ~\cite{imagej}.

\begin{figure}
\includegraphics{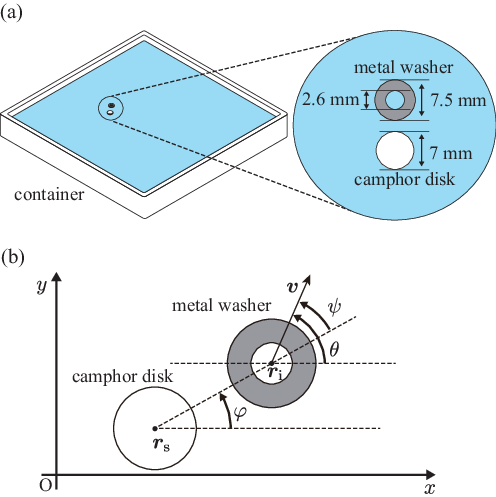}
\caption{(a) Experimental setup. A camphor disk and a metal washer were floated at the surface of the glycerol aqueous solution in a square-shaped container (245~mm $\times$ 245~mm $\times$ 25~mm). (b) Definition of the variables obtained from the experimental results.}
\label{fig1}
\end{figure}
After the metal washer and camphor disk were placed at the aqueous surface, they got close to each other owing to the attractive lateral capillary force. After a few seconds, they started a pairing-induced motion, in which the metal washer was in the front side and the camphor disk was in the rear side. Here, the time $t=0$ was defined as when the pairing-induced motion started. 
Figure~\ref{fig2} (a) shows the case with $C=0$ vol\%. The pair showed the circular motion, with either clockwise direction or counterclockwise direction chosen randomly. For low $C$ such a circular motion was typically observed. Figure~\ref{fig2} (b) shows, on the contrary, the straight motion for $C= 50.0$ vol\%. This type of straight  motion was observed for high $C$. In both cases, the pairs were bounced at the wall of the container. They changed direction at the collision, however, the circular/straight motion was quickly recovered once they departed from the wall. 

For the quantitative discussion on the pairing-induced motion, we measured the centers of mass (COMs) of the camphor disk $\bm{r}_{\rm{s}}^{j}$ and the metal washer $\bm{r}_{\rm{i}}^j$ at time $t=t^j$. First, we calculated the curvature of the trajectory to elucidate the transition between the circular and straight motions. It should be noted that the trajectories were represented by those of the washer. In detail, we obtained the velocity $\bm{v}^j=(\bm{r}_{\rm{i}}^{j+1} -\bm{r}_{\rm{i}}^j)/ \Delta t$, where $\Delta t = t^{j+1} - t^{j}=$$1/30~ \SI{}{s}$. Then, we obtained the angular velocity $\omega^j =(\theta^{j} - \theta^{j-1})/\Delta t$, where $\theta^j$ was calculated by the relation $\bm{v}^j =v^j (\cos \theta^j \bm{e}_x +\sin \theta^j \bm{e}_y )$, where $v^j > 0$. Here, $\bm{e}_x$ and $\bm{e}_y$ are unit vectors in the $x$- and $y$-directions, respectively. $\varphi^j$ is the angle of the vector of $\bm{r}_{\rm{i}}^{j}-\bm{r}_{\rm{s}}^{j}$ from the $x$-axis, and the drift angle is defined as $\psi^j=\theta^j -\varphi^j$. From $\omega^j$ and $v^j$, the unsigned curvature (the absolute value of the curvature) $\kappa^j=\kappa(t^j)$ was obtained as $\kappa^j = 2 \left| \omega^j \right| /(v^j +v^{j-1})$. 

The drift angle $\psi$ and angular velocity $\omega$ were fluctuated around a certain finite value during the circular motion (Fig.~\ref{fig2}(c)), while both were close to zero during the straight motion (Fig.~\ref{fig2}(d)). When the particles exhibited the circular motion, the drift angle $\psi$ had the opposite sign to the angular velocity $\omega$ (Fig.~\ref{fig2}(e)). It implies that the metal washer rotated at a smaller radius than the camphor disk. When the particles exhibited straight motion, the drift angle $\psi$ and angular velocity $\omega$ were almost zero (Fig.~\ref{fig2}(f)). 

Figures~\ref{fig_traj}(a)-(e) shows the trajectory for various $C$. Figures~\ref{fig_traj}(f)-(j) shows the distribution $p$ of the unsigned curvature $\kappa$ at each concentration corresponding to the trajectory shown in Figs.~\ref{fig_traj}(a)-(e). The distribution $p$ is normalized to be $\int p\rm{d}\kappa=1$. The distribution was taken only from the trajectories in the red box to prevent the effect of the sharp turns near the container walls. When $C$ was 0 vol\% and 12.5 vol\%, the distribution has a peak at $\kappa \simeq 0.03$ \SI{}{mm}$^{-1}$, indicating the trajectory is circular. In contrast, when $C$ was between 37.5 and 50.0 vol\%, the distribution exhibits a strong peak around $\kappa = 0$, which reflects a straight trajectory. In summary, these distributions shown in Figs.~{\ref{fig_traj}}(f)-(j) suggest that the circular trajectory changed into the straight one with an increase in $C$.

\begin{figure}
\includegraphics{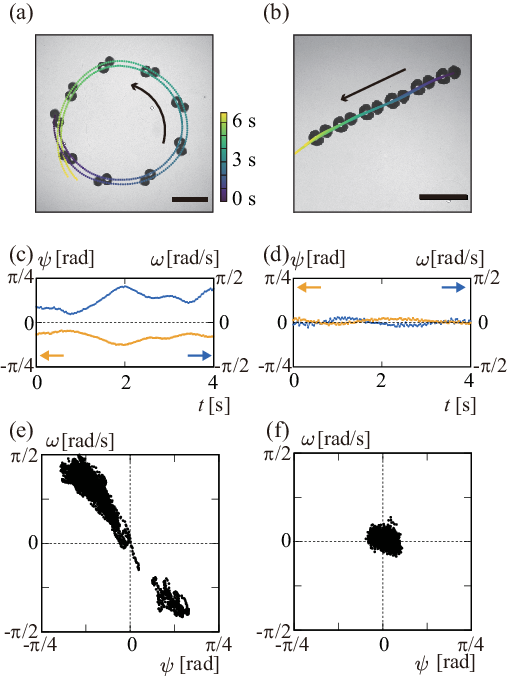}
\caption{(a, b) Superimposed images of the camphor disk and metal washer, whose trajectories are also denoted. Scale bar is \SI{30}{mm}. (a) Circular and (b) straight motions observed on the glycerol aqueous solutions at $C=$(a) 0 and (b) 50.0 \SI{}{vol\%}. (c,~d) Time series of the drift angle $\psi$ (orange) and angular velocity $\omega$ (blue) for (c) circular and (d) straight motions. The time $t$ corresponds to that in (a) and (b). (e,~f) Relationship between the drift angle $\psi$ and the angular velocity $\omega$ for (e) circular and (f) straight motions for the period from $t =0$ to 184 s. Corresponding videos are available in Supplemental Material~\cite{video}.}
\label{fig2}
\end{figure}

\begin{figure*}
\includegraphics{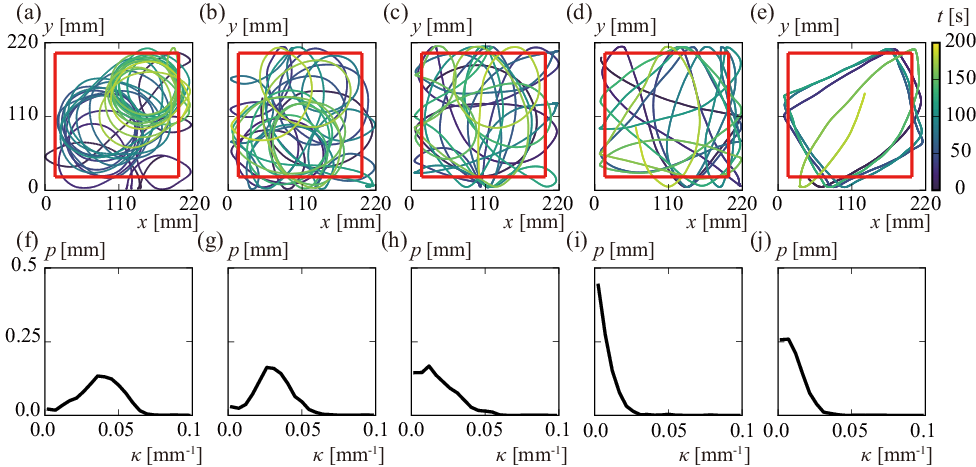}
\caption{(a-e) Trajectories of the pair represented by those of the washer. 
(f-j) Distribution $p$ of unsigned curvature $\kappa$ corresponding to (a)-(e).
The trajectories in the domain surrounded by red squares in (a)-(e) are used to obtain $p$.
The concentration of glycerol was $C$ was (a, f) 0, (b, g) 12.5, (c, h) 25.0, (d, i) 37.5 and (e, j) 50.0 \SI{}{vol\%}. The distribution $p$ shifts gradually to the lower $\kappa$ as $C$ increased, indicating that the circular motion gradually changed into the straight one.}
\label{fig_traj}
\end{figure*}

\section{Numerical simulations}
We construct a two-dimensional mathematical model to discuss the pairing-induced motion of the camphor disk and metal washer floating at the aqueous surface observed in the experimental system. In our model, the camphor disk and metal washer are regarded as the source and inert particles, respectively. We consider the time development of the source particle position $\bm{r}_{\rm{s}}(t)$, the inert particle position $\bm{r}_{\rm{i}}(t)$, and the concentration field $u(\bm{r},t)$ in the two-dimensional space, which corresponds to the aqueous surface. For simplicity, we consider that the inert particle does not affect the concentration field. We also neglect the hydrodynamic interaction due to the motion of the particles.

The dynamics for the concentration field is described as
\begin{equation}
\pd{u}{t}=D \Delta u-\alpha u+G(\bm{r},\bm{r}_{\rm{s}}).
\label{conc}
\end{equation}
Here, $D$ is the effective diffusion coefficient~\cite{kitahata_yoshinaga, kitahata_suematsu}, and $\alpha$ is the sublimation rate of the surface-active molecules. $G(\bm{r},\bm{r}_{\rm{s}})$ represents the supply rate of the molecules from the source particle located at $\bm{r}_{\rm{s}}$, 

\begin{equation}
G(\bm{r},\bm{r}_{\rm{s}}) = \cfrac{G_0}{S_0}H_\epsilon (\bm{r},\bm{r}_{\rm{s}},R_{\rm{s}}), 
\end{equation}
where $R_{\rm{s}}$ and $S_0$ denote the radius and area of the source particle, respectively. $G_0$ is the total supply rate of the surface-active molecules. $H_\epsilon$ is the smoothed step function defined as  
\begin{align}
H_\epsilon (\bm{r},\bm{r}',R')=\cfrac{1}{2}\cbk{1+\tanh{\cfrac{R'-|\bm{r}-\bm{r}'|}{\epsilon}}}.
\label{eq3}
\end{align}
Here, $\epsilon$ is a small positive parameter for smoothing. When $\epsilon \to +0$, the function $H_\epsilon (\bm{r}, \bm{r}', R')$ is replaced with a step function as
\begin{align}
H_0 (\bm{r},\bm{r}',R')= \left\{ \begin{array}{ll} 1,& |\bm{r}-\bm{r}'| \leq R' , \\ 0, &  |\bm{r}-\bm{r}'| > R'. \end{array}\right.
\end{align}

The dynamics for the motion of the source and inert particles are described as
\begin{align}
m_{\rm{s}} \odt{{\bm r}_{\rm{s}}}{t} =&-\eta_{\rm{s}} \od{{\bm r}_{\rm{s}}}{t} + {\bm{F}}_{\rm{conc}}\cbk{\gamma,{\bm r},{\bm r}_{\rm{s}}, R_{\rm{s}}}+{\bm{F}}_{\rm int} \cbk{{\bm r}_{\rm{i}}- {\bm r}_{\rm{s}}}, 
\label{eq4}
\end{align}
and
\begin{align}
m_{\rm{i}} \odt{{\bm r}_{\rm{i}}}{t} =&-\eta_{\rm{i}} \od{{\bm r}_{\rm{i}}}{t} + {\bm{F}}_{\rm{conc}}\cbk{\gamma,{\bm r},{\bm r}_{\rm{i}}, R_{\rm{i}}}+{\bm{F}}_{\rm int} \cbk{{\bm r}_{\rm{s}}- {\bm r}_{\rm{i}}}, 
\label{eq5}
\end{align}
where $m_{\rm{s}}$ and $\eta_{\rm{s}}$ denote the mass and viscous resistance coefficient for the source particle, respectively, while $m_{\rm{i}}$ and $\eta_{\rm{i}}$ denote the mass and viscous resistance coefficient for the inert particle, respectively.

The force originating from the surface tension gradient, $\bm{F}_{\rm{conc}}$, is defined as 
\begin{align}
\bm{F}_{\rm{conc}}\cbk{\gamma,\bm{r},\bm{r}',R'}=\iint_{\mathbb{R}^2}^{} [\bm{\nabla} \gamma] H_\epsilon (\bm{r},\bm{r}',R') d\bm{r},
\end{align}
which is illustrated in Fig.~\ref{fig3}(a). It should be noted that
\begin{align}
\iint_{\mathbb{R}^2}^{} [\bm{\nabla} \gamma] H_0 (\bm{r},\bm{r}',R') d\bm{r}  = \int_{\partial \Omega} \gamma \bm{n}(\bm{r}') dl',\label{da_to_dl}
\end{align}
holds when $\epsilon \to +0$. Here, $\partial \Omega$ is the periphery of the particle, and $dl'$ is a line element along it. $\bm{n}(\bm{r}')$ is a unit normal vector directing outward from the particle at the periphery. The expression in Eq.~\eqref{da_to_dl} shows that $\bm{F}_{\rm{conc}}$ represents the summation of the surface tension exerting perpendicular at the periphery of the particle. 

Here, we assume the following linear relationship between the surface tension $\gamma$ and concentration $u$:

\begin{equation}
\gamma=\gamma_0-\Gamma u,
\label{eq6}
\end{equation}
where $\gamma_0$ is the surface tension of the camphor-free aqueous phase. $\Gamma$ is a positive constant, reflecting that surface-active molecules decrease the surface tension of the aqueous phase~\cite{karasawa, kitahata_suematsu}.

The force $\bm{F}_{\rm{int}}$ is defined as 
\begin{equation}
\begin{split}
{\bm F}_{\rm int} \cbk{\bm{l}}
=F_\mathrm{int}( \left| \bm{l} \right| ) \frac{\bm{l}}{\left| \bm{l}\right|}=\left\{\begin{array}{ll}  f  \mathcal{K}_1 \cbk{q|{\bm{l}}|}\cfrac{{\bm{l}}}{|{\bm{l}}|}, & |\bm{l}|>R_{\rm{s}}+R_{\rm{i}},\\
(a{|{\bm{l}}|}+b)\cfrac{{\bm{l}}}{|{\bm{l}}|}, & |{\bm{l}}| \le R_{\rm{s}}+R_{\rm{i}},
\end{array}\right. \\
\end{split}
\label{cap_f}
\end{equation}
which describes the attractive lateral capillary force~\cite{l_cap_f} and short-range exclusive volume effect between the particles for $|\bm{l}| > R_{\rm{s}} + R_{\rm{i}}$ and $|\bm{l}| <R_{\rm{s}} + R_{\rm{i}}$, respectively (Fig.~\ref{fig3}(b)).
Here, $\bm{l}$ is the relative position vector from the considered particle to the other particle. $\mathcal{K}_1$ is the modified Bessel function of the second kind of order $1$, and $q$ is the inverse of the capillary length. The constant $a$ denotes an effective spring constant, which is related to the excluded volume. The constant $b$ is explicitly described as
\begin{equation}
b=f \mathcal{K}_1 (q(R_{\rm{s}}+R_{\rm{i}}))-(R_{\rm{s}}+R_{\rm{i}})a,
\end{equation}
so that the force shown in Eq.~\eqref{cap_f} becomes continuous. The surface tension modulation caused by camphor molecules, $\Gamma$, is negligible and thus we can neglect the dependence of the lateral capillary force on the concentration $u$. For simplicity, we set $m_{\rm{s}}=m_{\rm{i}}=m$, $\eta_{\rm{s}}=\eta_{\rm{i}}=\eta$, and  $R_{\rm s}=R_{\rm i}=R$.

In numerical calculations and theoretical analyses, we adopt the dimensionless form of the model. The dimensionless variables and coefficients are defined as
\begin{equation}
\begin{split}
&\tilde{t}=\alpha t,\quad\tilde{\bm{r}}=\sqrt{\cfrac{\alpha}{D}}\bm{r},\quad\tilde{q}=\sqrt{\cfrac{D}{\alpha}}q,\quad\tilde{\bm{l}}= \sqrt{\cfrac{\alpha}{D}}\bm{l},\\
&\tilde{G_0}=\cfrac{1}{\alpha S_0 u_0}G_0,\quad \tilde{\eta}=\cfrac{1}{m\alpha}\eta,\quad\tilde{\Gamma}=\cfrac{u_0}{m\alpha^2}\Gamma,\\
&\tilde{f}=\cfrac{1}{m\alpha\sqrt{D\alpha}}f, \quad \tilde{R}=\sqrt{\cfrac{\alpha}{D}}R, \quad\tilde{\epsilon}=\sqrt{\cfrac{\alpha}{D}}\epsilon,\\
&\tilde{a}=\cfrac{1}{m\alpha^2}a,\quad\tilde{b}=\cfrac{1}{m\alpha\sqrt{D\alpha}}b,\quad \tilde{u}=\cfrac{u}{u_0}.
\end{split}
\end{equation}
Here $u_0$ is the unit of concentration. The tildes ($\tilde{\ }$) are omitted hereafter for simplicity. The dimensionless forms are summarized as
\begin{align}
\pd{u}{t}=&  \Delta u- u+G_0 H_\epsilon (\bm{r},\bm{r}_{\rm{s}},R), \\
\odt{{\bm r}_{\rm{s}}}{t} =& -\eta \od{{\bm r}_{\rm{s}}}{t}- \Gamma \iint_{\mathbb{R}^2}^{} [\bm{\nabla}u ] H_\epsilon (\bm{r},\bm{r}_{\rm{s}},R) d\bm{r} \nonumber \\ 
&+ {\bm F}_{\rm int} \cbk{{\bm r}_{\rm{i}}- {\bm r}_{\rm{s}}}, \\ 
 \odt{{\bm r}_{\rm{i}}}{t} =& -\eta \od{{\bm r}_{\rm{i}}}{t}- \Gamma \iint_{\mathbb{R}^2}^{} [\bm{\nabla}u ]H_\epsilon (\bm{r},\bm{r}_{\rm{i}},R) d\bm{r} \nonumber \\
&+ {\bm F}_{\rm int} \cbk{{\bm r}_{\rm{s}}- {\bm r}_{\rm{i}}}, \\
H_\epsilon (\bm{r},\bm{r}',R')=& \cfrac{1}{2}\cbk{1+\tanh{\cfrac{R'-|\bm{r}-\bm{r}'|}{\epsilon}}}, \\
{\bm F}_{\rm int} \cbk{\bm{l}}= &\left\{\begin{array}{ll}  f  \mathcal{K}_1 \cbk{q|{\bm{l}}|}\cfrac{{\bm{l}}}{|{\bm{l}}|},&  |\bm{l}|>2R,\\
(a{|{\bm{l}}|}+b)\cfrac{{\bm{l}}}{|{\bm{l}}|},&  |{\bm{l}}| \le 2R, \\
\end{array}\right. \\
b=&f \mathcal{K}_1 (2Rq)-2Ra.
\end{align}

Numerical calculations were performed by changing $\eta$ and $\Gamma$ as parameters. $\Gamma$ represents the intensity of the driving force. The other parameters were $a=40$, $G_0=4/\pi$, $q=0.7$, $f=0.1$, $R=0.5$, and $\epsilon=0.1$. 

\begin{figure}
\includegraphics{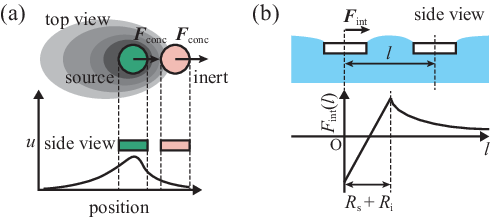}
\caption{Force applied to the particle. (a) Force originating from the concentration field. (b) Profile of attractive lateral capillary force and repulsive force owing to the short-range exclusive volume effect. The graph shows the force applied to the particle at the origin from another particle at $l=|\bm{l}|$ away.}
\label{fig3}
\end{figure}

The concentration field was calculated using the alternating direction implicit (ADI) method ~\cite{Numerical_Recipes_in_C}, and the positions of the source and inert particles were integrated using the Euler method. We adopted a periodic boundary condition to investigate the long-term behavior without the effect of the finite system size. 
The system size, time step, and spatial mesh were $L=25.6$, $\Delta t=0.005$, and $\Delta x=0.1$, respectively. The initial position of the inert particle was set at the center of the calculation area. The initial position of the source particle was set at the angles of $\pi/9$, $2\pi/9$, and $\pi/3$ from the $x$-axis with a distance of $1/q$ from the inert particle to check whether the effect of anisotropy of the spatial mesh configuration can be neglected. We confirmed such anisotropy of the spatial mesh is negligible at a steady state, but the statistics were taken with all of these initial configurations to eliminate possible differences. The initial velocity of the inert particle was $0$. The initial velocity of the source particle was set in the opposite direction to the inert particle with an absolute value of 0.01.

\begin{figure}
\includegraphics{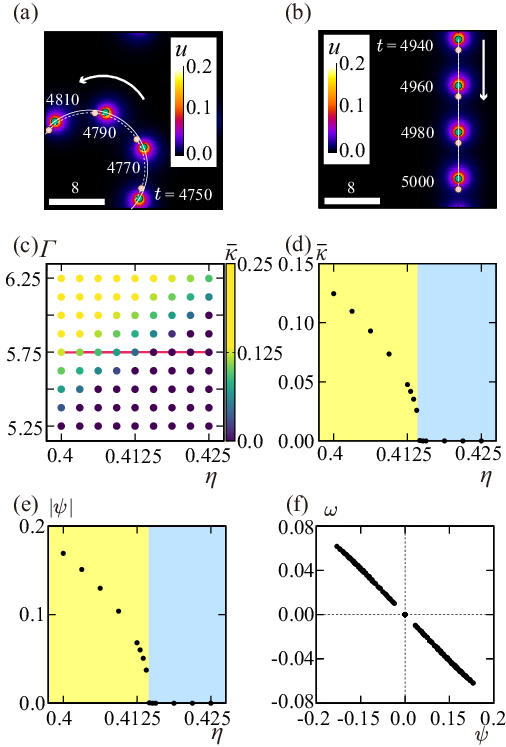}
\caption{(a, b) Superimposed images obtained using numerical calculations, in which the color map shows the profile of $u$. The green and pink disks represent the source and inert particles, respectively. The solid and dotted lines exhibit trajectories of the source and inert particles, respectively. The white arrows indicate the direction of motions. (a) Circular motion for $\Gamma= 5.75$ and $\eta = 0.4$. (b) Straight motion for $\Gamma= 5.75$ and $\eta = 0.425$. (c) Phase diagram that classifies circular and straight motions based on the mean value of the unsigned curvature, $\bar{\kappa}$, on the $\Gamma$-$\eta$ plane. The source and inert particles exhibited the circular motion for higher $\Gamma$ and lower $\eta$, while they exhibited the straight motion for lower $\Gamma$ and higher $\eta$. (d) Dependence of $\bar{\kappa}$ on $\eta$ for fixed $\Gamma=5.75$ on the red line in (c). (e) Dependence of $|\psi|$ on $\eta$ for fixed $\Gamma =5.75$ on the red line in (c). The yellow and light blue areas represent the parameter regions where circular and straight motions appeared, respectively in (d) and (e). (f) Relationship between the drift angle $\psi$ and angular velocity $\omega$.}
\label{fig4}
\end{figure}

The pair of the source and inert particles eventually exhibited steady motion, although the source particle alone did not exhibit self-propelled motion in the same parameter sets as shown in Fig.~\ref{1particle} in Appendix~\ref{1particle_ap}. The superimposed images obtained based on the numerical calculation are shown in Figs.~\ref{fig4}(a) and (b), in which the profile of the concentration field $u$ and the position of the source and inert particles are displayed. The trajectories of the source and inert particles are denoted as solid and dotted lines, respectively. They exhibited a straight motion for higher $\eta$ and lower $\Gamma$, and they exhibited a circular motion for smaller $\eta$ and larger $\Gamma$. In both the cases, the inert particle was in the front side, while the source particle was in the rear side. The configuration can be understood as follows. The inert particle escapes from the concentration field created by the source particle, while the attractive lateral capillary force keeps the relative distance between these two particles. We may also recognize the pair of the source and inert particles as a camphor boat whose relative positions are fixed~\cite{camphor_boat}. To discuss the transition between the straight and circular motions, the unsigned curvature of the trajectory of the source particle was evaluated. We confirmed that the unsigned curvature of the trajectory reached a steady value, and the values were almost the same independent of the initial condition for the particles configuration. Figure~\ref{fig4}(c) shows the phase diagram, in which the mean value of the unsigned curvature $\bar{\kappa}$ is shown. To obtain $\bar{\kappa}$, the time average from $t=225$ to $250$ was considered for each initial condition, and the mean value of the data  for three different initial conditions for the particle configuration was obtained. The smaller $\bar{\kappa}$ indicates the straight motion, while the larger $\bar{\kappa}$ indicates the circular motion. That is to say, when $\Gamma$ was larger and $\eta$ was smaller, they exhibited a circular motion. In contrast, when $\Gamma$ was smaller and $\eta$ was larger, they exhibited a straight motion. The transition shown in Fig.~\ref{fig4}(d) corresponds to the experimental observation in Figs.~{\ref{fig_traj}}(f)-(j). Figure~\ref{fig4}(e) shows the absolute value of the drift angle $\psi$. The transition of $|\psi|$ occurs at the same value of $\eta$ as that of $\bar{\kappa}$. When the source and inert particles exhibit circular motions, the drift angle $\psi$ possesses the opposite sign to the angular velocity $\omega$ (Fig.~\ref{fig4}(f)). It implies that the inert particle rotated with a smaller radius than the source particle. 

Figure~\ref{fig_vec} shows the forces acting on the source and inert particles based on the numerical calculation. 
For both particles, the net force, i.e., the summation of the 
viscous resistance forces ($\bm{F}^{\rm{s}}_{\rm{vis}}$, $\bm{F}^{\rm{i}}_{\rm{vis}}$), the lateral capillary forces ($\bm{F}^{\rm{s}}_{\rm{cap}}$, $\bm{F}^{\rm{i}}_{\rm{cap}}$), and the driving forces ($\bm{F}^{\rm{s}}_{\rm{conc}}$, $\bm{F}^{\rm{i}}_{\rm{conc}}$) by the concentration field, acts in the direction perpendicular to the particles velocities.
The net forces act as centripetal forces to keep them in a circular motion.
\begin{figure}
\includegraphics{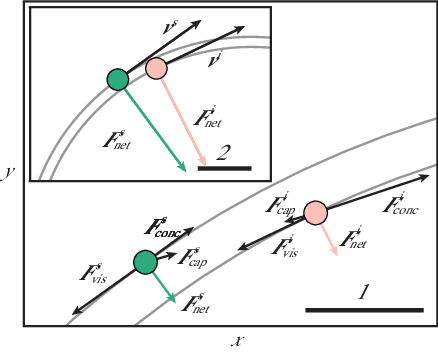}
\caption{Schematic illustration of forces acting on each particle in the circular motion. Arrows indicate the direction and the relative magnitude of these forces based on the numerically obtained data. The gray curves indicate the trajectories of the particles. The black arrows indicate the viscous resistance forces ($\bm{F}^{\rm{s}}_{\rm{vis}}$, $\bm{F}^{\rm{i}}_{\rm{vis}}$), the lateral capillary forces ($\bm{F}^{\rm{s}}_{\rm{cap}}$, $\bm{F}^{\rm{i}}_{\rm{cap}}$), and the driving forces ($\bm{F}^{\rm{s}}_{\rm{conc}}$, $\bm{F}^{\rm{i}}_{\rm{conc}}$) from the concentration field. The pink and green disks indicate the inert and source particles, respectively. The pink and green arrows show the net force acting on the inert and source particles. The scale of the net forces is enhanced by a factor of 5. The inset represents the wider view to illustrate the relation between trajectories and net forces. Here the scale of the net forces is further enhanced by a factor of 50.}
\label{fig_vec}
\end{figure}

We further evaluate the quantitative aspect of the numerical simulation. The diffusion length and the characteristic decay time are estimated to be $\sqrt{D/\alpha} \simeq 1 \times 10^{-2}$ \si{}{m} and $1/\alpha \simeq 1$ \si{}{s}, respectively, from the previous study~\cite{koyano_ocsi},  in which the motions of the camphor particle confined in a one-dimensional region was investigated. In this system, the camphor particle is reflected by the walls due to the concentration field and exhibits oscillatory motions. The orders of the diffusion length and the characteristic decay time are estimated from the distance of the reverse position from the wall and the period of the oscillation, respectively. In the numerical simulation, the particle and the system sizes were set as 1 and 25.6. Then we obtained the velocities of the order of 0.1. These values are estimated to be 10 mm, 256 mm, and 1 mm/s. In the experiment, the particle and system sizes were set as 7 mm and 245 mm, and the observed velocities were of the order of 1 to 10 mm/s. These values in the numerical calculations are consistent with those in the experiment.

We then evaluate the order of the force, starting from the estimation of the resistance coefficient. From the previous study~\cite{pozrikidis_2007}, the viscous resistance force $\bm{F}_{\rm{vis}}$ acting on a particle floating at an aqueous surface can be described as
\begin{equation}
	\bm{F}_{\rm{vis}} =-\eta \bm{v} \sim -a \pi \mu \bm{v},
	\label{res}
\end{equation}
Here, $a$ is the radius of the camphor particle, $\mu$ is the viscosity of the liquid phase, $\bm{v}$ is the velocity of the particles, and $\eta_{\rm e}$ is the resistance coefficient. The dimensionless resistance coefficient $\tilde{\eta}_{\rm e}$ is defined as
\begin{equation}
	\tilde{\eta}=\cfrac{\eta}{m \alpha} \sim \cfrac{a \pi \mu}{m \alpha},
\end{equation}
where $m$ is the mass of the particle and $\alpha$ is the sublimation rate of the camphor. In our experiment, $a=3.5$ \si{}{mm} and  $m=1\times10^{-4}$ \si{}{kg}, $\mu=1$ \si{mPa s}. Here  $\mu$ is estimated from the viscosity of pure water. Then, the dimensionless resistance coefficient based on the experimentally observed value is
\begin{equation}
	\tilde{\eta}_{\rm{e}}\simeq 6.3\times 10^{-1},
\end{equation}
which corresponds to $\eta_{\rm{e}} \sim 10^{-5}$ kg/s. 
In numerical simulation, we set $\tilde{\eta} \simeq 0.4$, which is consistent with $\tilde{\eta}_{\rm{e}}$.
Equations~\eqref{eq4},~\eqref{eq5} and~\eqref{res} lead the estimation of the driving force as,
\begin{equation}
	\bm{F}_{\rm{conc}} \simeq -\bm{F}_{\rm{vis}} = \eta \bm{v},
\end{equation}
when the particles are moving at a constant speed.
From the experimental results, the velocity of the pair is $|\bm{v}|\sim 10^{-2}$ \SI{}{m/s}. Then we estimate the driving force $|\bm{F}_{\rm{conc, e}}|\sim 10^{-6}$ N, which corresponds to $|\tilde{\bm{F}}_{\rm{conc, e}}| \sim 1$. In the numerical calculations, $|\bm{\tilde{v}}|\simeq 0.4 $ and hence $|\tilde{\bm{F}}_{\rm{conc}}|\sim 0.1$. These values in the numerical calculations are not too far from those in the experiment.

%\\

\section{Bifurcation analysis}

To elucidate the transition between the circular and straight pairing-induced motions of the source and inert particles, we perform the linear stability analysis for the straight motion at a constant velocity (Fig.~\ref{fig_anal}(a)) and discuss the transition based on the bifurcation theory. 

Considering that the linearity of the evolution equation in Eq.~\eqref{conc} except for the source term, the concentration field $u$ of the surface-active molecules that spread from the source particle can be described as a functional of the source particle position $\bm{r}_\textrm{s}(t)$. For the analyses, we adopt the approximation that the concentration field $u$ is represented as a function of the relative position $\bm{r} - \bm{r}_\mathrm{s}$ and the velocity of the source particle $\bm{v}_\mathrm{s} = d\bm{r}_\mathrm{s}/dt$. That is to say, the concentration field is represented as
\begin{align}
u(\bm{r}) = u_v(\bm{r} - \bm{r}_\mathrm{s}, \bm{v}_\mathrm{s}),
\end{align}
whose representative profile with $\bm{r}_\mathrm{s} = \bm{0}$ and $\bm{v}_\mathrm{s} = \bm{e}_x$ is shown in Fig.~\ref{fig_anal}(b). The exact solution of $u_v(\bm{r}, \bm{v})$ is obtained in the form of infinite series, whose explicit expression together with the brief derivation is shown in Appendix.
It is noted that previous studies\cite{golestanian, Rotator, 3Dprinter} often introduce the interaction through the concentration field by considering the one depending only on the source particle position, while our model consider the concentration field depending on the source particle velocity as well as the position.
This concentration field exerts the force on the source or inert particle with a radius of $R$ located at $\bm{r}'$ as
\begin{align}
\bm{F} =& -\Gamma \iint_{\mathbb{R}^2} \left[\nabla u_v \left( \bm{r} - \bm{r}_\mathrm{s} , \bm{v}_\mathrm{s} \right)\right] H_0(\bm{r}, \bm{r}', R) d\bm{r} \nonumber \\
 =& -\Gamma \iint_{\mathbb{R}^2} \left[\nabla u_v \left( \bm{r}, \bm{v}_\mathrm{s} \right) \right] H_0(\bm{r}, \bm{r}' - \bm{r}_\mathrm{s}, R) d\bm{r} \nonumber \\
\equiv& \bm{F}_v( \bm{r}'-\bm{r}_\mathrm{s}, \bm{v}_\mathrm{s}, R) .
\end{align}
Thus, the force working on the particle through the concentration field is represented by $\bm{F}_v$, which is a function of the relative position of the focused particle from the source particle and the velocity of the source particle. Owing to this approximation, our original system comprised the partial differential equation (PDE) for the concentration field and ordinary differential equations (ODEs) for the particle position is simplified into the system described by the two second-order ODEs for $\bm{r}_\mathrm{s}$ and $\bm{r}_\mathrm{i}$. 

We discuss the linear stability of the straight pairing-induced motion. Our system is now expressed as an autonomous dynamical system with the eight degrees of freedom, i.e., $\bm{r}_{\rm{s}}$, $\bm{r}_{\rm{i}}$, $\bm{v}_{\rm{s}} = d\bm{r}_{\rm{s}}/dt$, and $\bm{v}_{\rm{i}} = d\bm{r}_{\rm{i}}/dt$. For the analysis, we introduce the position of the COM $\bm{r} = \left(\bm{r}_\mathrm{s} + \bm{r}_\mathrm{i} \right)/2$ and the relative position $\bm{\ell} = \bm{r}_\mathrm{i} - \bm{r}_\mathrm{s}$ in the place of $\bm{r}_\mathrm{s}$ and $\bm{r}_\mathrm{i}$. The velocity of the COM and relative velocity are introduced as $\bm{v} = \left( \bm{v}_\mathrm{s} + \bm{v}_\mathrm{i} \right) /2$ and $\bm{w} = \bm{v}_\mathrm{i} - \bm{v}_\mathrm{s}$, respectively. Our simplified dynamical system based on $\bm{r}$, $\bm{\ell}$, $\bm{v}$, and $\bm{w}$ is explicitly described as
\begin{align}
\od{\bm{r}}{t} =& \bm{v}, \label{eq_r}
\end{align}
\begin{align}
\od{\bm{v}}{t} =& -\eta \bm{v} + \frac{1}{2} \left[ \bm{F}_v\left(\bm{0}, \bm{v} - \frac{1}{2}\bm{w}, R\right) \right. \nonumber \\
&\qquad \left. + \bm{F}_v \left( \bm{\ell} , \bm{v} - \frac{1}{2} \bm{w} ,R \right)  \right] , \label{eq_v}
\end{align}
\begin{align}
\od{\bm{\ell}}{t} =& \bm{w}, \label{eq_ell}
\end{align}
\begin{align}
\od{\bm{w}}{t} =& - \eta \bm{w} - \bm{F}_v\left(\bm{0}, \bm{v} - \frac{1}{2} \bm{w} ,R\right) \nonumber \\
& \qquad + \bm{F}_v\left( \bm{\ell} , \bm{v} - \frac{1}{2} \bm{w} ,R \right) - 2 \bm{F}_\mathrm{int}(\bm{\ell}) . \label{eq_w}
\end{align}

First, we construct the solution corresponding to the straight pairing-induced motion in the positive $x$ direction. Considering that the inert particle precedes in the straight pairing-induced motion, as shown in Fig.~\ref{fig_anal}(a), the solution is described using a constant speed $v_0$ and a constant interval $\ell_0$ as
\begin{align}
\bm{r}(t) =  v_0 t \bm{e}_x,
\end{align}
\begin{align}
\bm{v}(t) = v_0 \bm{e}_x,
\end{align}  
\begin{align}
\bm{\ell}(t) = \ell_0 \bm{e}_x,
\end{align} 
\begin{align}
\bm{w}(t) = \bm{0},
\end{align} 
The constant speed $v_0$ and $\ell_0$ should be determined, so that the two relations related to the force balance in the $x$ direction,
\begin{align}
-2\eta v_0 \bm{e}_x + \bm{F}_v(\bm{0}, v_0 \bm{e}_x ,R) + \bm{F}_v(\ell_0 \bm{e}_x ,v_0 \bm{e}_x ,R) = \bm{0}, \label{anal_vel}
\end{align}
and
\begin{align}
 -\bm{F}_v(\bm{0}, v_0 \bm{e}_x ,R) + \bm{F}_v(\ell_0 \bm{e}_x ,v_0 \bm{e}_x ,R) - 2\bm{F}_\mathrm{int}(\ell_0 \bm{e}_x ) = \bm{0}, \label{anal_ell}
\end{align}
should hold.

The perturbation for the linear stability analysis is described as
\begin{align}
\bm{r}(t) = v_0 t \bm{e}_x + \delta \bm{r}(t) = v_0 t \bm{e}_x + \delta r_x(t)\bm{e}_x + \delta r_y(t)\bm{e}_y,
\end{align}  
\begin{align}
\bm{v}(t) = v_0 \bm{e}_x + \delta \bm{v}(t)  = v_0 \bm{e}_x + \delta v_x(t)\bm{e}_x + \delta v_y(t)\bm{e}_y,
\end{align}
\begin{align}
\bm{\ell}(t) &= \ell_0 \bm{e}_x + \delta \bm{\ell}(t) = \ell_0 \bm{e}_x+ \delta \ell_x(t)\bm{e}_x + \delta \ell_y(t)\bm{e}_y, 
\end{align}
\begin{align}
\bm{w}(t) &= \delta \bm{w} (t) =  \delta w_x(t)\bm{e}_x + \delta w_y(t)\bm{e}_y.
\end{align}
The linearized equations are separated into two independent parts
\begin{align}
\frac{d}{dt}  \left( \begin{array}{c} \delta \ell_x \\ \delta v_x \\ \delta w_x \end{array}\right) = A_x \left( \begin{array}{c} \delta \ell_x \\ \delta v_x \\ \delta w_x \end{array}\right) , \label{eq_deltavx}
\end{align}
\begin{align}
\frac{d}{dt} \left( \begin{array}{c} \delta \ell_y \\ \delta v_y \\ \delta w_y \end{array}\right) &=  A_y  \left( \begin{array}{c} \delta \ell_y \\ \delta v_y \\ \delta \omega_y \end{array}\right),
\end{align}
and a slave equation,
\begin{align}
\frac{d}{dt} \delta \bm{r} = \delta \bm{v},
\end{align}
where $A_i$ $(i = x,y)$ is explicitly described as
\begin{align}
A_i =  \left( \begin{array}{ccc} 0 & 0 & 1 \\ \frac{f_i}{2} & -\eta + \frac{g_i + h_i}{2} & - \frac{g_i + h_i}{4} \\ f_i  - 2p_i & g_i - h_i & -\eta - \frac{g_i - h_i}{2} \end{array}\right) .
\end{align}
To obtain the aforementioned linearized equation, we used the relation in Eqs.~\eqref{anal_vel} and \eqref{anal_ell}.
Here, $f_i$, $g_i$, $h_i$, and $p_i$ are defined as
\begin{align}
f_i =  \lim_{\varepsilon \to 0} \frac{1}{\varepsilon} \left[ \bm{F}_v (\ell_0 \bm{e}_x + \varepsilon \bm{e}_i, v_0\bm{e}_x , R) - \bm{F}_v (\ell_0 \bm{e}_x, v_0 \bm{e}_x , R)\right] \cdot \bm{e}_i, \label{eq_fi}
\end{align}
\begin{align}
g_i =  \lim_{\varepsilon \to 0} \frac{1}{\varepsilon} \left[ \bm{F}_v (\ell_0 \bm{e}_x , v_0 \bm{e}_x + \varepsilon \bm{e}_i , R) - \bm{F}_v (\ell_0 \bm{e}_x, v_0 \bm{e}_x , R)\right] \cdot \bm{e}_i,
\end{align}
\begin{align}
h_i =  \lim_{\varepsilon \to 0} \frac{1}{\varepsilon} \left[ \bm{F}_v (\bm{0} , v_0 \bm{e}_x + \varepsilon \bm{e}_i , R) - \bm{F}_v (\bm{0}, v_0 \bm{e}_x , R)\right] \cdot \bm{e}_i,
\end{align}
\begin{align}
p_i =  \lim_{\varepsilon \to 0} \frac{1}{\varepsilon} \left[ \bm{F}_\mathrm{int} (\ell_0 \bm{e}_x + \varepsilon \bm{e}_i) - \bm{F}_\mathrm{int} (\ell_0 \bm{e}_x ) \right] \cdot \bm{e}_i. \label{eq_pi}
\end{align}
Eigenvalues are obtained as the solution of the characteristic polynomial of $A_i$,
\begin{align}
&\lambda^3 + (2\eta - h_i)\lambda^2 + \left( \eta^2 - h_i \eta + 2p_i - f_i\right) \lambda \nonumber \\
& \qquad + \left(2 p_i - f_i \right)\eta - (g_i + h_i) p_i + f_i h_i  = 0. \label{eig}
\end{align}
Here, we consider the isotropy of the system, i.e., the solution in which the straight pairing-induced motion in any direction should exist. Therefore, Eqs.~\eqref{anal_vel} and Eqs~\eqref{anal_ell} hold if we substitute $v_0(\cos \nu \bm{e}_x + \sin \nu \bm{e}_y)$ and $\ell_0(\cos \nu \bm{e}_x + \sin \nu \bm{e}_y)$ for $v_0 \bm{e}_x$ and $\ell_0 \bm{e}_x$, respectively, where $\nu$ is a small parameter. Up to the first order of $\nu$, we obtain
\begin{align}
(2 p_y - f_y) \eta -p_y (g_y + h_y) + f_y h_y  = 0.
\end{align}
Therefore, the characteristic polynomial for $A_y$ is simplified as
\begin{align}
\lambda \left[ \lambda^2 + (2\eta - h_y)\lambda + \eta^2 - h_y \eta + 2p_y - f_y \right]  = 0. \label{eig2}
\end{align}
That is to say, the matrix $A_y$ possesses a zero eigenvalue, and the corresponding eigenvector is ${}^\mathrm{t}(\ell_0, v_0,0 )
$.

\begin{figure}
\includegraphics{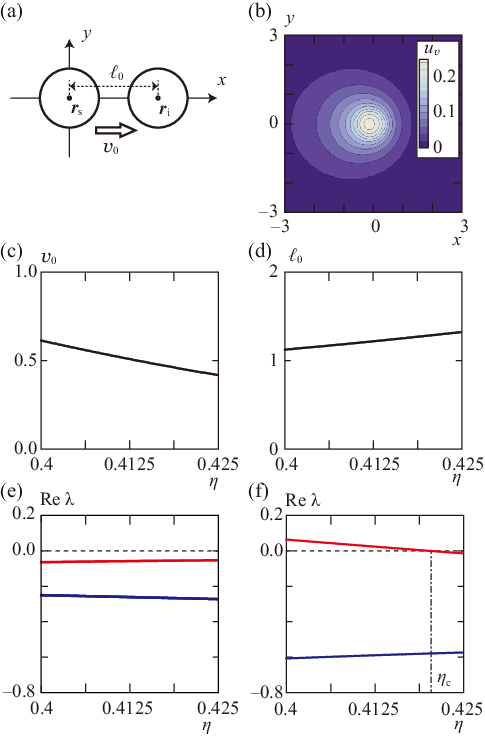}
\caption{(a) Coordinate system and straight pairing-induced motion. (b) Representative profile of $u_v(\bm{r}, \bm{v}_\mathrm{s})$, where $\bm{v}_\mathrm{s} = \bm{e}_x$. The contours at every interval of 0.2 are shown.  (c) $v_0$ against $\eta$ for $\Gamma = 5.75$. (d) $\ell_0$ against $\eta$ for $\Gamma = 5.75$. (e) Real parts of eigenvalues $\lambda$ of $A_x$ calculated for each $\eta$. The eigenvalue around $\mathrm{Re}\ \lambda \simeq -0.05$ (plotted with red) is real, while other eigenvalues around $\mathrm{Re}\,\lambda \simeq -0.25$ (plotted with blue) are complex. (f) Real parts of the eigenvalues $\lambda$ of $A_y$ calculated for each $\eta$. The zero eigenvalue is not plotted.}
\label{fig_anal}
\end{figure}

Using the aforementioned equations, we investigated the linear stability by adopting parameter values that were used in the numerical calculation. In the numerical evaluation, we truncated the infinite series of  $u_v$ in Eq.~\eqref{eq16} till $n = 10$. The limits of $\varepsilon$ in Eqs.~\eqref{eq_fi}--\eqref{eq_pi} were calculated by setting $\varepsilon = 10^{-3}$. For both approximations, we confirmed that the accuracy was within $10^{-3}$. 

For the evaluation, we first numerically obtained $v_0$ and $\ell_0$ for $\eta$ and $\Gamma$ based on Eqs.~\eqref{anal_vel} and \eqref{anal_ell}. We used the values which satisfied the equations with the accuracy of $10^{-3}$. The obtained values of $v_0$ are plotted against $\eta$ with constant $\Gamma = 5.75$ in Fig.~\ref{fig_anal}(c), and those for $\ell_0$ in Fig.~\ref{fig_anal}(d). $v_0$ decreased with an increase in $\eta$, while $\ell_0$ increased with an increase in $\eta$.
 
Using the obtained values of $v_0$ and $\ell_0$, we calculated $f_i$, $g_i$, $h_i$, and $p_i$ $(i = x,y)$, and then calculated the eigenvalues of $A_x$ and $A_y$ using Eqs.~\eqref{eig} and \eqref{eig2}, respectively. The real parts of the eigenvalues of $A_x$ are plotted against $\eta$ in Fig.~\ref{fig_anal}(e), while those of $A_y$ are plotted against $\eta$ in Fig.~\ref{fig_anal}(f). It should be noted that the zero eigenvalue is not shown in Fig.~\ref{fig_anal}(f). As shown in Fig.~\ref{fig_anal}(e), the straight pairing-induced motion with a translational speed $v_0$ at a distance of $\ell_0$ between the two particle is stable as far as the perturbation in the $x$-axis direction is concerned. As shown in Fig.~\ref{fig_anal}(f), the maximum real parts of the eigenvalues of $A_y$ changed its sign at $\eta = \eta_\mathrm{c} \simeq 0.42$. This means that the straight pairing-induced motion is stable for $\eta > \eta_\mathrm{c}$ and unstable for $\eta < \eta_\mathrm{c}$. This result qualitatively agrees with numerical results, though the threshold value is slightly greater in the theoretical analysis. Considering that the eigenvalue whose real part changes its sign at $\eta = \eta_\mathrm{c}$ does not have an imaginary part, and that the system is symmetric with the $x$-axis, the transition between the circular and straight pairing-induced motions is most likely classified as a pitchfork bifurcation. As higher-order terms are not calculated, we cannot distinguish whether the bifurcation is supercritical or subcritical. Based on the numerical results shown in Figs.~\ref{fig4} (d) and (e), the bifurcation is classified into a supercritical pitchfork bifurcation.

The transition from the straight to the circular motion can be intuitively understood as follows. The net force acting on the pair is created by the force due to the concentration field because the attractive lateral capillary force satisfies the action-reaction law. In the limit of the low velocity, the concentration field is radially symmetric around the source particle. Then, the net force acting on the the inert particle always directs along the source to the inert particle. As a result, the pair move straight. Such a situation becomes unstable when the velocity of the pair exceeds a threshold predicted by the linear stability analysis.

It should be noted that the linear stability analysis can be performed if we adopt the point-source approximation. We confirmed that the qualitatively same bifurcation structure is reproduced using the point-source approximation, though the bifurcation point is seriously different from the numerical results, which are quantitatively consistent with experimental results.

\section{Conclusion}
In this study, we focus on the pairing-induced motion of the source and inert particles. In the experiments, we used the camphor disk and metal washer as the source and inert particles, respectively. After we floated them on glycerol aqueous solution, they attracted each other through the attractive lateral capillary force. Then, they showed circular and straight motions on the aqueous solution with lower and higher glycerol concentration, respectively. We constructed a mathematical model to discuss the transition between the circular and straight motions. In the model, we considered time developments for the positions of the source particle, inert particle, and concentration field formed by the source particle. In numerical calculations, we reproduced circular and straight motions corresponding to the experimental results. Furthermore, we performed the linear stability analysis based on the straight pairing-induced motion and the obtained results were quantitatively consistent with the numerical results. The analysis suggested that the transition can be understood in terms of the pitchfork bifurcation. 

In the current analysis, the force originating from the concentration fields, $\bm{F}_v$, is approximated by taking up to the first-order terms with respect to time derivative of the source particle position. This expression of the force is exact when the source particle moves at a uniform velocity, and it enabled us to discuss the transition between the straight and circular motions. Indeed, we did not observe such transition when the concentration field is represented as the function of the relative position as mentioned in the previous study~\cite{golestanian}.
We expect that further complex dynamics such as zig-zag, quasi-periodic, and chaotic motions~\cite{tarama} can be realized by adequately choosing experimental conditions and/or numerical parameters. To address such complicated motions based on the bifurcation analysis, we need to improve our analysis method. One possible candidate is to include higher-order terms in the expansion of the concentration field with respect to the velocity, acceleration, jerk, and combination of these terms~\cite{koyano,koyano_Gryciuk,koyano_suematsu}. 

We should bear in mind that the effective interaction induced by concentration fields is non-reciprocal, i.e., it breaks the action-reaction law. Experimentally, a single camphor disk shows self-propulsion in the absence of a washer. However, numerically and theoretically, a pair of source and inert particles can have self-propulsion even in the parameter range where a single source particle cannot have self-propulsion. The pair-induced motion is a direct consequence of the non-reciprocal interaction, and our study describes the relevance of its dynamic aspect. The well-known camphor-water system should be re-investigated based on the new perspective of non-reciprocal interaction~\cite{cira_nature,Zarzar_Nature_Chem,Fruchart_Nature}. 

One can easily conceive the extension of the present system to one with multiple particles. The collective motion of particles driven by the dynamics of the concentration field has not yet been understood in detail. The knowledge drawn from our study can be applied to develop experimental and numerical systems. The expansion of the concentration field as used in the present study can be adopted to construct simple models with multiple particles by extracting the essential dynamics of the concentration field. We believe that the possibility of extracting complex motions from such a simple system will lead to a better understanding of the collective motions of migrating cells and bacteria.

\begin{acknowledgments}
The authors acknowledge Professor Jerzy Gorecki (Polish Academy of Sciences), Professor Hiroaki Ito (Chiba University), and Professor Masaharu Nagayama (Hokkaido University) for their fruitful discussion.
This study was supported by JSPS KAKENHI Grant Nos.~JP19J00365, JP19H05403, JP20K14370, JP20H02712, JP21H00409, JP21H00996, and JP21H01004 and the Cooperative Research Program of ``Network Joint Research Center for Materials and Devices: Dynamic Alliance for Open Innovation Bridging Human, Environment and Materials'' (Nos.~20211014 and 20214004).
Furthermore, this study was supported by JSPS and PAN under the Japan-Poland Research Cooperative Program (No.~JPJSBP120204602) and by JST, the establishment of University fellowships towards the creation of
science technology innovation (No.~JPMJFS2107).
\end{acknowledgments}

\appendix

\section{The motion of the source particle}\label{1particle_ap}
Figure~\ref{1particle}(a) shows the phase diagram that represents the steady velocity of the source particle $v_{\rm{s}}$ on the $\Gamma$-$\eta$ plane. Here, $v_{\rm{s}}$ is defined as the mean velocity from $t = 225$ to $250$. Figure~\ref{1particle}(b) shows the dependence of $v_{\rm{s}}$ on $\eta$ for fixed $\gamma=5.75$ represented by the red line in Fig.~\ref{1particle}(a). Both the phase and the bifurcation diagrams indicate that the source particle did not exhibit self-propulsion in the parameter sets corresponding to those in Fig.~\ref{fig4}(c).

\begin{figure}
\includegraphics{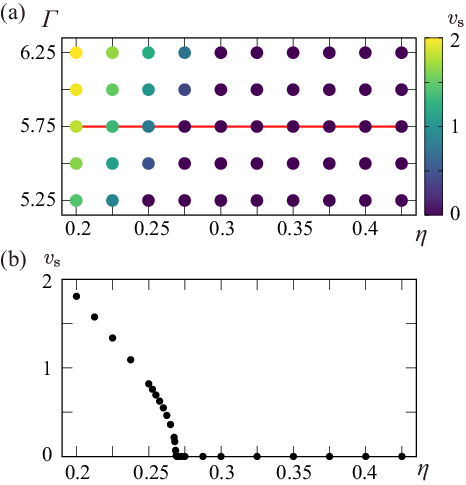}
\caption{(a) Phase diagram that represents the single source particle motion based on the steady velocity $v_s$. The source particle exhibited self-propelled motion for the higher $\Gamma$ and the lower $\eta$, while it stopped for the lower $\Gamma$ and the higher $\eta$. (b) Dependence of  $v_{\rm{s}}$ on $\eta$ for fixed $\Gamma$ = 5.75 indicated by the red line in (a).}
\label{1particle}
\end{figure}

\section{Derivation of $u_v$}
In this Appendix, we derive the concentration field $u_v(\bm{r}, \bm{v})$ generated by the source particle whose position and velocity are $\bm{0}$ and $\bm{v}$, respectively. We consider the situation that the source particle moves at a constant velocity $\bm{V} = V \bm{e}_x$. We introduce the co-moving frame $\bm{\rho}$, in which the source particle is located at the origin. The dynamics for the concentration field is described as
\begin{align}
\frac{\partial u}{\partial t} - V \bm{e}_x \cdot \nabla u = \Delta u - u + G_0 H_0 (\bm{\rho}, \bm{0}, R).
\end{align}
The stationary solution $U(\bm{\rho}; V)$ should satisfy
\begin{align}
 - V \bm{e}_x \cdot \nabla U = \Delta U - U + G_0 H_0 (\bm{\rho}, \bm{0}, R). \label{eq_stationary}
\end{align}
The general solution of the homogeneous equation for Eq.~\eqref{eq_stationary}
\begin{align}
 - V \bm{e}_x \cdot \nabla U = \Delta U - U,
\end{align}
is given as the linear combination of
\begin{align}
U(\rho, \phi; V) = \mathcal{I}_n \left(k \rho \right) \exp \left( - \rho \mu \cos\phi \right) \exp \left(i n\phi \right),
\end{align}
and
\begin{align}
U(\rho, \phi; V) = \mathcal{K}_n \left(k \rho \right) \exp \left( - \rho \mu \cos\phi \right) \exp \left(i n\phi\right),
\end{align}
in the polar coordinates $\rho$ and $\phi$, which satisfy $\bm{\rho} = \rho (\cos\phi \bm{e}_x + \sin\phi \bm{e}_y )$~\cite{kitahata_koyano_book}.  Here, $n$ is an integer, $k=\sqrt{1 + V^2 / 4}$, and $\mu=V/2$. $\mathcal{I}_n$ and $\mathcal{K}_n$ are the modified Bessel functions of the first and second kinds of order $n$, respectively.
Therefore, the solution of Eq.~\eqref{eq_stationary} is obtained as
\begin{align}
   U(\bm{\rho}; V)
    &= \left\{ \begin{array}{ll} \displaystyle{ G_0 - G_0 \sum_{n=0}^\infty \alpha_n  \mathcal{I}_n \left(k \rho \right) \exp \left( - \rho \mu \cos\phi \right) \cos n\phi},  \\
						\hspace{\fill} \rho \leq R, \\
						\displaystyle{ G_0 \sum_{n=0}^\infty \beta_n  \mathcal{K}_n \left(k \rho \right) \exp \left( - \rho \mu \cos\phi \right) \cos n\phi},  \\
						\hspace{\fill} \rho > R.  \end{array}  \right. 
\label{eq16}
\end{align}
Here, the coefficients $\alpha_n$ and $\beta_n$ are obtained from the continuity condition of $U$ and $\nabla U$ at the periphery of the particle. They are explicitly given as
\begin{align}
    \alpha_n =& \zeta_n \left[- k R \mathcal{I}_n \left(\mu R \right) \mathcal{K}'_n \left(k R \right)  + \mu R \mathcal{I}'_n \left(\mu R \right) \mathcal{K}_n \left(k R \right) \right],
\label{eq17}
\end{align}
\begin{align}
    \beta_n =& \zeta_n \left[ k R \mathcal{I}_n \left(\mu R \right) \mathcal{I}'_n \left( k R \right) -\mu R \mathcal{I}'_n \left(\mu R \right) \mathcal{I}_n \left(k R \right) \right],
\label{eq18}
\end{align}
where
\begin{align}
    \zeta_n = \left\{ \begin{array}{ll} 1, & n = 0, \\ 2, & n \geq 1, \end{array} \right.
\label{eq19}
\end{align}
and the prime (${}'$) denotes the derivative. When the velocity of the source particle is in the arbitrary direction, the concentration field $u_v$ is obtained as
\begin{align}
u_v(\bm{r}, \bm{v}) = U( \mathcal{R}(-\vartheta) \bm{r}; \left|\bm{v} \right|), 
\end{align}
where $\vartheta$ holds $\bm{v} = \left|\bm{v} \right| (\cos\vartheta \bm{e}_x + \sin\vartheta\bm{e}_y )$, where $\mathcal{R}(\vartheta)$ is the rotation matrix
\begin{align}
\mathcal{R}(\vartheta)= \left( \begin{array}{cc} \cos \vartheta & -\sin \vartheta \\ \sin \vartheta & \cos \vartheta \\ \end{array}\right).
\end{align}

\end{document}